\documentclass[a4paper,11pt]{article} 
\usepackage[dvips]{graphics} 
\usepackage[british]{babel} 
\usepackage{amssymb,amsmath,bm} 
\usepackage[ansinew]{inputenc} 
\usepackage{epsfig} 
\usepackage{natbib}
\begin{document} 
	\begin{center} 
		\Large\textbf{Introducing the Q-based interpretation of quantum theory}\vspace{0.5cm}\\
\normalsize Simon Friederich\vspace{0.5cm} 	\end{center}\vspace{0.5cm}
		\small{\textbf{Abstract:} This article outlines a novel interpretation of quantum theory: the Q-based interpretation. The core idea underlying this interpretation, recently suggested for quantum field theories by \citet{drummondreid}, is to interpret the phase space function $Q$---a transform of the better known Wigner function---as a proper probability distribution, roughly analogous to the probability distribution $\rho$ in classical statistical mechanics.
		
	 Here I motivate the Q-based interpretation, investigate whether it is empirically adequate, and outline some of its key conceptual features. I argue that the Q-based interpretation is attractive in that it promises having no measurement problem, is conceptually parsimonious and has the potential to apply elegantly to relativistic and field-theoretic contexts.}

\tableofcontents

\section{Introduction}

This article outlines key features of a novel interpretation of quantum theory. I propose to call it the Q-based interpretation because it attributes a central role to a function on phase space known as the Q-function. The core idea behind the Q-based interpretation was recently suggested for quantum field theories by \citet{drummondreid}. That idea is simply to interpret the Q-function $Q(\mathbf q,\mathbf p)$ as a proper probability distribution on phase space, analogous to the probability distribution $\rho(\mathbf q,\mathbf p)$ in classical statistical mechanics.

The aims of this article are: to motivate the Q-based interpretation, to investigate whether it is empirically adequate, and to outline some of its key conceptual features. The Q-based interpretation is attractive in various respects, notably, because it promises having no measurement problem, because it is conceptually parsimonious in that it does not add any new elements to the formalism of quantum theory, and because it applies to relativistic and field-theoretic contexts.

The structure of the following sections is as follows: Section 2 provides a sketch of phase space distribution functions in quantum mechanics and reviews some formal properties of the Q-function, which is one such distribution. Section 3 introduces the basic idea of the Q-based interpretation---interpreting the Q-function as a proper probability distribution over phase space---and reviews some results by \citet{drummond} concerning microdynamics of fields that would give rise to such statistics in bosonic quantum field theory. Sections 4 and 5 outline prima facie difficulties for interpreting the Q-function as a proper probability distribution and show how one can hope to overcome them when explicitly considering the dynamics of measurement processes. Sections 6 and 7 explain how the Q-based interpretation, when combined with an epistemic account of the quantum state, avoids various no-go theorems by rejecting a temporal locality assumption that \citet{leiferpusey} call $\lambda$-mediation. I also consider the claim by \citet{drummondreid} that the Q-based interpretation entails retrocausality and argue that there is no compelling argument for it. Section 8 turns to the prospects for applying the Q-based interpretation to relativistic quantum theories and quantum field theories. Finally, section 9 concludes the paper with a brief summary of virtues of the Q-based interpretation and an outlook at issues that deserve further investigation.

\section{Phase Space Distribution Functions in Quantum Mechanics}

Quantum mechanics might not suffer from the measurement problem if it could be interpreted as a probabilistic theory on phase space analogous to classical statistical mechanics. Notably, its empirical success might not be mysterious if quantum expectation values $\langle \mathcal A\rangle_{\rm quantum}$ of dynamical variables $\mathcal A(\mathbf q,\mathbf p)$ could be understood as classical phase space averages in terms of some function $f(\mathbf q,\mathbf p)$:
\begin{equation}
\langle \mathcal A\rangle_{\rm quantum}=\int \mathcal A(\mathbf q,\mathbf p) f(\mathbf q,\mathbf p)d\mathbf q d\mathbf p\,,\label{probdens}
\end{equation}
where $f(\mathbf q,\mathbf p)$ is a probability density. For that to be possible, the integral of $f(\mathbf q,\mathbf p)$ over phase space would have to be $1$, and $f(\mathbf q,\mathbf p)$ would have to be non-negative everywhere. In that case, an interpretation of $f(\mathbf q,\mathbf p)$ as expressing partial information about the location of the quantum system in phase space might be available, analogously to how $\rho(\mathbf q,\mathbf p)$ in classical statistical mechanics is usually interpreted as expressing partial information about phase space location.

Intriguingly, identities similar to Eq.\ (\ref{probdens}) hold for a variety of phase space functions $F(\mathbf q,\mathbf p)$, namely, identities of the form:
\begin{equation}
\langle \hat A(\mathbf {\hat q},\mathbf {\hat p})\rangle_{\rm quantum}=\int A(\mathbf q,\mathbf p) F(\mathbf q,\mathbf p)d\mathbf q d\mathbf p\,.\label{quasiprobabilitydensities}
\end{equation}
The function $A(\mathbf q,\mathbf p)$ that appears on the right-hand side here is obtained from the operator $\hat A(\hat{\mathbf q},\hat{\mathbf p})$ which appears on the left-hand side by replacing the position and momentum operators $\hat{\mathbf q}$ and $\hat{\mathbf p}$ with $n$-tuples of ordinary real numbers. For any dynamical variable $\mathcal A(\mathbf q,\mathbf p)$ of interest we may insert for $\hat A(\hat{\mathbf q},\hat{\mathbf p})$ the  Hilbert space linear operator onto which it is mapped by quantization.

Eq.\ (\ref{quasiprobabilitydensities}) is not quite the same as Eq.\ (\ref{probdens}) with $\mathcal A(\mathbf q,\mathbf p)$ promoted to $\hat A(\hat{\mathbf q},\hat{\mathbf p})$ because $F(\mathbf q,\mathbf p)$ may not have the formal properties of a probability density and because $A(\mathbf q,\mathbf p)$ in Eq.\ (\ref{quasiprobabilitydensities}) may differ from $\mathcal A(\mathbf q,\mathbf p)$ in eq.\ (\ref{probdens}) due to the non-commutability of $\hat{\mathbf q}$ and $\hat{\mathbf p}$. This will become relevant below.

In quantum mechanics it is always possible to express any $\langle \hat A(\mathbf {\hat q},\mathbf {\hat p})\rangle_{\rm quantum}$ in the form Eq.\ (\ref{quasiprobabilitydensities}). However, there is no single phase space function $F(\mathbf q,\mathbf p)$ for which it holds uniformly. Rather, the function $F(\mathbf q,\mathbf p)$ for which Eq.\ (\ref{quasiprobabilitydensities}) obtains depends on the ordering of operators in $\hat A(\mathbf {\hat q},\mathbf {\hat p})$. For operators $\hat A(\mathbf {\hat q},\mathbf {\hat p})$ that are symmetrized in terms of the order in which $\hat{\mathbf q}$ and $\hat{\mathbf p}$ appear (e.\ g. the operator $1/2(\hat{\mathbf q}\hat{\mathbf p}+\hat{\mathbf p}\hat{\mathbf q})$), this is the so-called Wigner function; for standard-/antistandard ordered operators ($\hat{\mathbf q}$ before/after $\hat{\mathbf p}$ in all product terms), these are the so-called Kirkwood functions; for normal-ordered operators (creation before annihilation operators, see Eqs.\ (\ref{annihilation}) and (\ref{creation}) below for the definition of these operators) the so-called Glauber-Sudarshan function; and for antinormal-ordered operators (annihilation before creation operators) the so-called Husimi function.

These phase space functions $F(\mathbf q,\mathbf p)$ are convenient calculational tools. According to \citet{lee}, who articulates a standard perspective, they cannot be anything more than that. The Heisenberg uncertainty principle makes it impossible to regard them as proper probability functions:
\begin{quote}
	It has been realized from the early days (\citet{wigner}) that there is no unique way of defining a quantum phase-space distribution function. The concept of a joint probability at a $(q,p)$ phase-space point is not allowed in quantum mechanics due to the Heisenberg uncertainty principle. The quantum phase-space distribution function should therefore be considered as simply a mathematical tool that facilitates quantum calculations, and as such one can devise any `quasiprobability' distribution function that one wishes as long as it yields a correct description of physically observable quantities. In many situations the Wigner distribution function does a respectable job, and yet there are cases where distribution functions that have different properties than the Wigner distribution function are called for. Other distribution functions that have been considered in the past include those of Glauber-Sudarshan (\citet{glauber} [...], \citet{sudarshan})) \citet{husimi} and \citet{kirkwood}. (\citet{lee}, p.\ 150) 
\end{quote}
However, prima facie at least, Lee's claim that the Heisenberg principle rules out `[t]he concept of a joint probability at a $(q,p)$ phase-space point' is not compelling: if we had an interpretation that included probabilities ascribed to sharp phase space locations for all quantum systems at all times and understood the Heisenberg principle as applying only on an aggregate level of measurements, there might be no measurement problem. Such an interpretation, far from ruled out from the start, would be very welcome! But there are other, better, reasons to doubt that there can be such an interpretation. Those reasons have to do with two issues: (i) some functions $F(\mathbf q,\mathbf p)$ do not have the formal features of a probability density and, as already mentioned, (ii) the dynamical variable $\mathcal A(\mathbf q,\mathbf p)$ in Eq. (\ref{probdens}) and the function $A(\mathbf q,\mathbf p)$ in Eq.\ (\ref{quasiprobabilitydensities}) may differ. I elaborate on issue (i) now and come back to issue (ii) in Section 4.

The most well-known phase space distribution function $F(\mathbf q,\mathbf p)=W(\mathbf q,\mathbf p)$, the  Wigner function, illustrates issue (i): as mentioned, if the operator $\hat A(\hat{\mathbf q},\hat{\mathbf p)}$ is symmetric with respect to $\hat{\mathbf q}$ and $\hat{\mathbf{p}}$, then using $W(\mathbf q,\mathbf p)$ for $F(\mathbf q,\mathbf p)$ fulfils Eq.\ (\ref{quasiprobabilitydensities}). However, $W(\mathbf q,\mathbf p)$ is typically negative somewhere on phase space, and this makes it impossible to interpret it as a proper probability distribution. The same holds for some other phase space distributions $F(\mathbf q,\mathbf p)$.

The phase space regions where the Wigner function is negative are `small', however, in that they do not span more than a few intervals of $\hbar$ in any direction. As a consequence, if $W(\mathbf q, \mathbf p)$ is smoothed by means of a convolution with a minimum-uncertainty Gaussian wave packet (a so-called Weierstrass transformation), one obtains the Husimi function, $F^H(\mathbf q, \mathbf p)$, which is positive semi-definite everywhere and has all of the formal properties of a probability density. For a particle in one dimension, it has the form:
\begin{eqnarray}
F^H(q,p)=\frac{2}{\pi}\int dq'dp'\exp\left(-\frac{m\kappa(q'-q)^2}{\hbar}-\frac{(p'-p)^2}{\hbar m\kappa}\right)W(q',p')\,,
\end{eqnarray}
where the right hand side specifies the convolution operation. As mentioned above, $F^H(q, p)$ fulfils Eq.\ (\ref{quasiprobabilitydensities}) if $\hat A(\hat{\mathbf q},\hat{\mathbf p})$ is anti-normally ordered, i.e. with $\hat{\mathbf q}$ and $\hat{\mathbf p}$ expressed in terms of creation and annihilation operators (see Eqs.\ (\ref{creation}) and (\ref{annihilation})) and the latter to the left of the former.

For the harmonic oscillator, there is a conceptually preferred value of $\kappa$, namely, the oscillation frequency $\omega$. If one sets $\kappa$ to $\omega$ in the Husimi function, the function thereby obtained is called the Q-function. For a particle in one dimension it can be written as
\begin{eqnarray}
Q(\mathbf q,\mathbf p)=\frac{1}{\pi}\langle\alpha_{q, p}|\hat\rho|\alpha_{q, p}\rangle\,.\label{Qdef1}
\end{eqnarray}
Here $\hat\rho$ is the quantum state in density operator form and $|\alpha_{q,p}\rangle$ the coherent state wave packet centred around $(q, p)$. In position space, it is given by:
\begin{eqnarray}
\alpha_{q, p}(x)=\left(\frac{m\omega}{\pi\hbar}\right)^{1/4}\exp\left(-m\omega(x-q)^2/(2\hbar)-ipx/\hbar\right)\,.\label{coherentstate}
\end{eqnarray}
The coherent states $|\alpha_{q, p}\rangle$ are characterized by the fact that they are eigenstates of the components of the annihilation operator $\hat a$:
\begin{eqnarray}
\hat a|\alpha_{q,p}\rangle=\alpha_{q, p}|\alpha_{q, p}\rangle.
\end{eqnarray}
The annihilation operator and its adjoint, the creation operator $\hat a^\dagger$, are defined by:
\begin{eqnarray}
\hat a= \frac{1}{\sqrt{2\hbar m\omega}}\left(m\omega\hat q+i\hat p\right),\label{annihilation}\\
\hat a^\dagger= \frac{1}{\sqrt{2\hbar m\omega}}\left(m\omega\hat q-i\hat p\right).\label{creation}
\end{eqnarray}
Phase space can be parametrized either by position and momentum variables $q$ and $p$ or, equivalently, by the complex variable $\alpha$ that is obtained when replacing the position and momentum operators in Eq.\ (\ref{annihilation}) by position and momentum phase space variables.

All formulas Eqs.\ (\ref{coherentstate}) -- (\ref{creation}) tailored to the harmonic oscillator can be generalized to bosonic quantum field theory, where $\hbar\omega$ is the energy of a free particle. There, in a lattice approximation with $N$ degrees of freedom, Eq.\ (\ref{Qdef1}) generalizes to
\begin{eqnarray}
Q(\boldsymbol \alpha)=\frac{1}{\pi^N}\langle\boldsymbol\alpha|\hat\rho|\boldsymbol\alpha\rangle\,.\label{Qdef}
\end{eqnarray}
Here again the variable $\boldsymbol\alpha$ parametrizes the full phase space of the (field) theory.

The Q-function---like the Husimi function, of which it is a special case---has the formal properties of a probability density in that it is normalized to $1$ and positive semi-definite. (For an abstract general characterization of Q-functions, which also applies to fermionic quantum field theories, see (\citep{Rosales-Zarate}), as reviewed below in Section 8.) When substituted for the function $F$ in Eq.\ (\ref{quasiprobabilitydensities}), the Q-function provides the expectation value of $\mathcal A$ if the operator $\hat A(\mathbf q,\mathbf p)$ is anti-normally ordered, i.e. has annihilation operators to the left of creation operators in all products of such operators.

The dynamics of the Q-function follow directly from Eq.\ (\ref{Qdef}) together with the von Neumann time-evolution equation for $\hat\rho_t$:
\begin{eqnarray}
i\hbar\frac{d\hat\rho_t}{dt}=[\hat H,\hat\rho_t]\,,\label{vonNeumann}
\end{eqnarray}
namely,
\begin{eqnarray}
\frac{dQ(\boldsymbol\alpha,t)}{dt}=-\frac{i}{\pi\hbar}Tr\big\lbrace[\hat H,\hat\rho_t]|\boldsymbol\alpha\rangle\langle\boldsymbol\alpha|\big\rbrace\,.\label{vonNeumannQ}
\end{eqnarray}
\citet{drummond} investigates which form this time-evolution equation Eq.\ (\ref{vonNeumann}) takes in bosonic quantum field theory with a generic (up to) quartic Hamiltonian. This approach does not carry over to generic first-quantized Hamiltonians as in non-relativistic quantum mechanics. As Drummond shows, for phase space coordinates $\phi$ that are appropriately chosen linear combinations of rescaled versions of the real and imaginary parts of the $\alpha_i$, time-evolution of the Q-function is governed by a diffusion (Fokker-Planck) equation
\begin{eqnarray}
\frac{dQ(\phi,t)}{dt}=\frac{\partial}{\partial\phi^\mu}\big\lbrack-A^\mu(\phi)+\frac{1}{2}\frac{\partial}{\partial\phi^\mu}D^\mu(\phi)\big\rbrack Q(\phi,t)\,. \label{diffusion}
\end{eqnarray}
The coordinates $\phi$ can be chosen such that the diffusion matrix, which is traceless, is diagonal, with $D^\mu\ge0$ for half of all degrees of freedom $\mu$ and $D^\mu\le0$ for the other half. Thus, intuitively, time-evolution of the Q-function in bosonic quantum field theory with a quartic Hamiltonian corresponds to diffusion in positive and negative time directions, differentially for different degrees of freedom. For practical purposes, though, it is often easier to determine the time-evolution of $Q$ by using the von Neumann equation Eq.\ (\ref{vonNeumann}) for $\hat\rho_t$ and evaluating it using Eq.\ (\ref{Qdef}).

\section{What Is the Q-based Interpretation?}
The core idea of the Q-based interpretation is that any quantum system has a determinate location in phase space at all times and that the Q-function must be interpreted as a proper probability density over phase space. Using traditional terms, this makes the Q-based interpretation a `hidden variables' interpretation in the same sense in which de Broglie-Bohm theory is one. It aims to solve the measurement problem by denying that a quantum state is a complete description of a quantum system and postulates determinate values for all dynamical variables at all times.

According to the Q-based interpretation, a quantum system's probability of being in some region $\Delta$ of phase space is given by the integral of $Q(\mathbf q,\mathbf p)$ over $\Delta$:
\begin{equation}
Pr[\mathcal A\in\Delta]=\int_{\mathbf q,\mathbf p|_{\mathcal A\in\Delta}} Q(\mathbf q,\mathbf p)d\mathbf q d\mathbf p\,.\label{q_born}\end{equation}
Thus the correct expectation values to ascribe to dynamical variables $\mathcal A(\mathbf q,\mathbf p)$ are obtained from Eq.\ (\ref{probdens}) when replacing $f(\mathbf q,\mathbf p)$ by $Q(\mathbf q,\mathbf p)$:
\begin{equation}
\langle \mathcal A\rangle_{\rm phys}=\int \mathcal A(\mathbf q,\mathbf p) Q(\mathbf q,\mathbf p)d\mathbf q d\mathbf p\,.\label{q_int}
\end{equation}

One may argue that the Q-based interpretation can provide a fully-fledged solution to the measurement problem only if it comes with an account of the underlying microdynamics that give rise to Q-function dynamics, in analogy to the role played by the guidance equation in de Broglie-Bohm theory. Are there basic dynamical principles that create such aggregate behaviour as encoded in Eqs.\ (\ref{q_born}) and (\ref{q_int})?

\citet{drummond} provides a positive answer to this question for bosonic quantum field theory where time-evolution of the Q-function has the form Eq.\ (\ref{diffusion}). As he shows, such aggregate behaviour arises from stochastic field trajectories whose probabilities are given by real-valued path integrals over the exponential of a suitably chosen time-symmetric action. This requires a little unpacking.

Consider a bosonic field $\phi$ and denote its values at two different times $t_f>t_0$ by $\phi_0$ and $\phi_f$. As explained above, in Q-function dynamics as characterized by Eq.\ (\ref{diffusion}), half of the field degrees of freedom are characterized by `forward-in-time' diffusion and the other half by `backward-in-time' diffusion, depending on the sign of $D^\mu$. As \citet{drummond} shows, under these circumstances the conditional probability $P(\phi_{x,f},\phi_{y,0}|\phi_{x,0},\phi_{y,f})$, which mixes initial and final field configurations for the different degrees of freedom and aligns their order with the direction of diffusion, is proportional to a real path integral
\begin{eqnarray}
P(\phi_{x,f},\phi_{y,0}|\phi_{x,0},\phi_{y,f})\sim\int\mathcal D\phi\exp\left(-\int_{t_0}^{t_f}\mathcal L(\phi,\dot{\phi})dt\right)\,.\label{path}
\end{eqnarray}
Here $\mathcal L(\phi,\dot{\phi})$ is a so-called `central difference' Langrangian that treats $\phi_x$ and $\phi_y$ differently, see (\citet{drummond}, Eqs.\ (110)-(112)) for details. Drummond develops a scheme for solving Eq.\ (\ref{path}) based on an extra dimension (Sect.\ IV) and demonstrates solutions for examples (Sect.\ V). Note that, because the integrand in Eq.\ (\ref{path}) is real and non-negative, this path integral allows precisely the `ignorance interpretation' over paths (in this case, field histories) that is unavailable for the standard Feynman path integral, which has a complex integrand.\footnote{The Q-based interpretation may thus fulfil the hope for a field-based `all-at-once' interpretation of the path integral articulated in (\citet{wharton}).}

Drummond and Reid interpret Eq.\ (\ref{path}) as encoding retrocausality because the later field configuration $\phi_{y,f}$ seemingly undergoes diffusion into the earlier one $\phi_{y,0}$. I argue below (Section 7), however, that this is not a compelling reason for a retrocausal interpretation.

The time-mixing nature of Eq.\ (\ref{path}) may offer us a clue about another interesting question namely, why the wave function $\psi$ is widely regarded as central in quantum theory rather than $Q$, which is puzzling from the point of view of the Q-based interpretation.\footnote{I would like to thank an anonymous referee for highlighting this feature of the Q-based interpretation.} The reason may have to do with the fact that, for the purposes of reasoning about physical problems, we are temporally localized creatures. It comes natural to us and is often dictated by practical considerations to impose single-time boundary conditions on functions such as $Q$ or $\psi$ and consider time-evolution towards the future. The Schr\"{o}dinger equation can be treated in this manner: it is typically quite convenient to impose some initial condition $\psi_0$ for the wave function and evolve it in time. However, when we impose an initial condition $Q(\phi_0)$ onto the Q-function, Eq.\ (\ref{path}) is of no immediate help because $\phi_{x,0}$ and $\phi_{y,0}$ appear on opposite sides of the `$|$'. This may be a (small) part of the reason why the Q-function has so far been regarded as of secondary importance compared with the wave function even though, according to the Q-based interpretation, the reverse perspective is more adequate.

\section{Can the Q-based Interpretation Possibly Be Empirically Viable?}
Having put forward the idea of interpreting the Q-function as a proper probability density, it is time to consider what is perhaps the most natural worry about it, namely, that it is not empirically viable. Notably, one may worry that Eq.\ (\ref{q_born}) is incompatible with how probabilities are normally derived in standard quantum mechanics as based on Weyl quantization. This brings us back to issue (ii) that arises when trying to interpret a phase space function as a proper probability distribution, as announced in Section 2. One instance of Eq.\ (\ref{quasiprobabilitydensities}) is
\begin{equation}
\langle \hat A(\mathbf {\hat q},\mathbf {\hat p})\rangle_{\rm quantum}=\int A(\mathbf q,\mathbf p) Q(\mathbf q,\mathbf p)d\mathbf q d\mathbf p\,,\label{Q_antinorm}
\end{equation}
which holds if $\hat A(\hat{\mathbf q},\hat{\mathbf p})$ is in anti-normal order and where $A(\mathbf q,\mathbf p)$ is obtained from $\hat A(\mathbf {\hat q},\mathbf {\hat p})$ by replacing operators with numbers. For the Q-based interpretation to be empirically viable, this $A(\mathbf q,\mathbf p)$ must equal the original dynamical variable $\mathcal A(\mathbf q,\mathbf p)$ as subjected to quantization. And these two, in turn, are uniformly identical only if the chosen quantization procedure promotes dynamical variables to Hermitian linear operators that are anti-normally ordered, i.e. with annihilation before creation operators. This is the case in so-called Berezin (or anti-Wick) quantization, which differs from the more commonly used Weyl quantization, where dynamical variables are promoted to operators that are symmetrized in terms of the $\hat q_i$ and $\hat p_i$.\footnote{See (\citet{landsman}, pp.\ 460-1) for a concise review and comparison.}

The fact that symmetrized operators are not in general in anti-normal order leads to the above issue (ii) in that, for symmetrized $\hat A(\mathbf {\hat q},\mathbf {\hat p})$, Eq.\ (\ref{Q_antinorm}) may not be fulfilled. The quantum expectation values for dynamical variables promoted to operators according to Weyl quantization will generally differ from the physical expectation values according to the Q-based interpretation as computed per Eq.\ (\ref{q_int}). But the well-established empirical success of quantum mechanics, one may worry, requires using Weyl quantization for mapping dynamical variables onto operators. The Q-based interpretation, in contrast is empirically adequate only if Berezin quantization is.

To give an example, the discrepancy between Weyl and Berezin quantization can be illustrated with the dynamical variable `square of position', $q^2$, for a particle in one dimension. In terms of the phase space variable
\begin{eqnarray}
\alpha=\frac{1}{\sqrt{2\hbar m\omega}}\left(m\omega q+i p\right)
\end{eqnarray}
we obtain
\begin{eqnarray}
q^2=\frac{\hbar}{2m\omega}\left(\alpha^2+\alpha^{*2}+2\alpha\alpha^*\right)\,.
\end{eqnarray}
Berezin quantization promotes this to the anti-normally ordered
\begin{eqnarray}
\frac{\hbar}{2m\omega}\left(\hat a^2+\hat a^{\dagger\,2}+2\hat a\hat a^\dagger\right)\,.\label{Berezin_operator}
\end{eqnarray}
Weyl quantization, in contrast, promotes $q^2$ to the operator $\hat q^2$. To compare $\hat q^2$ with Eq.\ (\ref{Berezin_operator}), we must bring $\hat q^2$ into anti-normal order:
\begin{eqnarray}
\hat q^2=\frac{\hbar}{2m\omega}\left(\hat a^2+\hat a^{\dagger\,2}+2\hat a\hat a^\dagger-1\right)\,.\label{Weyl_operator}
\end{eqnarray}
Due to the last term in Eq.\ (\ref{Weyl_operator}) this differs from Eq.\ (\ref{Berezin_operator}).

The quantum mechanical expectation value of $q^2$ expressed in terms of the Q-function can be computed based on Eqs.\ (\ref{Q_antinorm}) and (\ref{Weyl_operator}):
\begin{eqnarray}
\langle q ^2\rangle_{\rm quantum}=\int dqdp\ q^2\ Q(q,p) -\frac{\hbar}{2m\omega}\,.
\end{eqnarray} 
This differs by $-\frac{\hbar}{2m\omega}$ from the expectation value according to Berezin quantization, which, according to Eqs.\ (\ref{Q_antinorm}) and (\ref{Berezin_operator}), is
\begin{eqnarray}
\langle q ^2\rangle_{\rm quantum}=\int dqdp\ q^2\ Q(q,p)\,.
\end{eqnarray}
By Eq.\ (\ref{q_int}), this is also the physical expectation value according to the Q-based interpretation. Analogous discrepancies between quantum mechanics based on Weyl quantization on the one hand and the Q-based interpretation on the other are obtained for all dynamical variables where Weyl quantization promotes dynamical variables to operators that are not in anti-normal order.

A defender of the Q-based interpretation may respond by suggesting that associating dynamical variables differently with Hermitian linear operators than is usually done might not alter the empirical content of the theory.\footnote{I would like to thank an anonymous referee for suggesting this.} The operator that is usually taken to represent the dynamical variable $q^2$ does yield a correct expectation value, though not for $q^2$ but for $q^2-\frac{\hbar}{2m\omega}$. The revised association between dynamical variables and operators in the Q-based intepretation may seem unnatural, but it is actually a harmless theory-internal re-labelling procedure and no reason to doubt the empirical adequacy of that interpretation.

This response may not convince everyone. One may object that we often know quite well the value of which dynamical variable any given measurement device determines and that we know quite well the results of which measurements we must compare with the outcomes of which calculations. To give a simple macroscopic example, we may know that some traffic-monitoring device which measures vehicle velocity $v$ really does measure $v$ rather than, say, $v-10\,km/h$. If we have a traffic-predicting theory $T_1$ that predicts $v$ correctly for all observed vehicles and another one, $T_2$, that predicts the same numbers as $T_1$, but as values of $v-10\,km/h$ rather than of $v$, then $T_1$ is empirically adequate, whereas $T_2$ is not. There are important differences between this example and the question of empirical adequacy of the Q-based interpretation---notably, the suggested re-labelling of operators in terms of dynamical variables by switching to Berezin quantization does not merely concen one single dynamical variable measured by one specific device---but we may take it as a warning that we should not take the empirical adequacy of the Q-based interpretation for granted.

For another `wrong' result delivered by the Q-based interpretation recall that, according to textbook quantum mechanics, for a pure state $|\psi\rangle$, the probability density and the probability current density in space are
\begin{eqnarray}
\rho_\psi(\mathbf q)&=&|\psi(\mathbf q)|^2\,,\\
\mathbf j_\psi(\mathbf q)&=&|\psi(\mathbf q)|^2\ S(\mathbf q)\,,
\end{eqnarray}
where $\psi(\mathbf q)$ is the quantum state evaluated at the position $\mathbf q$, and $S(\mathbf q)$ is the angle of the wave function in the polar representation: \[\psi(\mathbf q)=|\psi(\mathbf q)|\exp(iS(\mathbf q)/\hbar).\] From the Wigner function, these quantities are obtained by integrating out the momentum degree(s) of freedom (dropping the time argument for simplicity):
\begin{eqnarray}
\rho_\psi(\mathbf q)&=&\int W(\mathbf q,\mathbf p)d\mathbf p\,,\label{wignerrho}\\
\mathbf j_\psi(\mathbf q)&=&\int \mathbf p\ W(\mathbf q,\mathbf p)d\mathbf p\,.\label{wignerj}
\end{eqnarray}
But these identities do not hold if we replace the Wigner function by the Q-function. In that case, the corresponding integrals read (see (\citet{colomes}, Eqs.\ (14), (15) and Appendix A), considering the one-dimensional case for simplicity)
\begin{eqnarray}
\rho_Q(q)&=&\int Q(q,p)dp\label{marginal1}\\ &&=\sqrt{\frac{m\omega}{2\hbar}}\int\exp\left(-m\kappa(q-q')^2/\hbar\right)|\psi(q')|^2dq'\,,\nonumber\\
j_Q(q)&=&\int p\;Q(q,p)dp =\label{marginal2}\\
&&\sqrt{\frac{m\omega}{2\hbar}}\int\exp\left(-m\kappa(q-q')^2/\hbar\right)|\psi(q')|^2\frac{\partial S(q')}{\partial q'}dq' \,.\nonumber
\end{eqnarray}
These quantities are in general not identical to $\rho_\psi(q)$ and $j_\psi(q)$, respectively. But the latter seem to be empirically accessible. Worryingly, interpreting the Q-function as a proper probability distribution over phase space does not allow one to recover them as weighted phase space averages.

As a sidenote, however, one may observe that $\rho_Q$ and $\mathbf j_Q$ at least conform to a continuity equation, which expresses the conservation of probability in space under the Q-based interpretation:
\begin{eqnarray}
\frac{\partial\rho_Q(\mathbf q,t)}{\partial t}=-div\ \mathbf{j}_Q(\mathbf q,t)\,.\label{continuity}
\end{eqnarray}
One can derive this equation by starting from the well-known quantum mechanical continuity equation
\begin{eqnarray}
\frac{\partial\rho_\psi(\mathbf q,t)}{\partial t}=-div\ \mathbf{j_\psi}(\mathbf q,t)\,,\label{standardcontinuity}
\end{eqnarray}
expressing $\rho_\psi$ and $j_\psi$ in terms of the Wigner function using Eqs.\ (\ref{wignerrho}) and (\ref{wignerj}), performing a Weierstrass transform on both sides of Eq.\ (\ref{standardcontinuity}) and using the fact that the Weierstrass transformation commutes with the temporal and spatial derivatives if the Wigner function is sufficiently smooth. The fact that Eq.\ (\ref{continuity}) holds signals that interpreting the Q-function as a proper probability distribution is at least internally coherent.

Summing up, one may think that the idea of interpreting of the Q-function as a proper probability function is a non-starter: quantum mechanical expectation values of dynamical variables as usually computed are not always given by Q-function-weighted classical phase space integrals, and the marginal distributions of the Q-function do not correspond to the  quantum probability density and probability current density. 

However, as proponents of de Broglie-Bohm theory often highlight, our epistemic access to the values of dynamical variables of microsystems is indirect and depends on our knowledge of the configurations of macroscopic objects that we use as pointers or displays of measurement apparatuses. Bell puts it concisely (and a leading Everettian concurs (\citet{wallace}, p. 21)):
\begin{quote}
	[I]n physics the only observations we must consider are position observations, if only the positions of instrument pointers. It is a great merit of the de Broglie-Bohm picture to force us to consider this fact. If you make axioms, rather than definitions and theorems, about the `measurement' of anything else, then you commit redundancy and risk inconsistency. (\citet{bell}, p.\ 166) 
\end{quote}
Thus, to check whether an interpretation of the Q-function as a proper probability distribution over phase space might be empirically adequate, what we really have to check is whether it produces the right probabilities for what Bell calls `the positions of instrument pointers'. The positions of instrument pointers over time are encoded in their phase space locations, thus we have to check whether interpreting the Q-function as a proper probability distribution delivers the same (for all practical purposes) predictions for the phase space locations of macroscopic objects in `measurement' contexts as standard quantun mechanics does.

Let us model the measurement apparatus $A$ as a macroscopic harmonic oscillator with $N$ degrees of freedom, $N\gg1$. Distinct `pointer settings' correspond to non-overlapping phase space regions $\Gamma_i$ of $A$'s phase space that have macroscopic dimensions. Suppose that the measurement of the system $S$ that we are performing is one that we usually characterize as `projective with respect to the non-degenerate basis $\lbrace|B_i\rangle\rbrace$'. Suppose further that, at the beginning of the measurement process, the quantum state that we assign to $S$ is a generic superposition
\begin{eqnarray}
\sum_ic_i|B_i\rangle,
\end{eqnarray}
of the basis states $|B_i\rangle$. The standard view of quantum mechanics is that the possible measurement outcomes are the eigenvalues $B_i$ of the operator $\hat B$ corresponding to the dynamical variable that is `measured' and that these outcomes are realized with probabilities $|c_i|^2$, respectively. But the eigenvalues are not directly accessible by observation, only the pointer settings are. Hence, to indicate that the Q-based interpretation is empirically adequate, one must show that---provided that the measurement interaction has led to a suitable association between $S$ and $A$---the probability $P(j)$ of the pointer setting being in $\Gamma_j$ at the end of measurement is (to a high degree of approximation) given by $|c_j|^2$ in that, if $\boldsymbol\alpha$ is the phase space variable for the apparatus $A$,
\begin{eqnarray}
P(j)&=&\int_{\boldsymbol\alpha\in\Gamma_j} d\boldsymbol\alpha\ Q_{red,A}(\boldsymbol\alpha)\nonumber\\
&\approx&|c_j|^2\,.\label{announce}
\end{eqnarray} 
I will now show that this holds, given reasonable assumptions about the measurement process.

Consider first a situation where the state assigned to $S$ prior to the measurement interaction is one of the eigenstates $|B_j\rangle$ of $\hat B$. To accept the apparatus $A$ as functioning properly, one will require that the post-measurement pointer setting is in a specific phase space region $\Gamma_j$ with probability 1. (One would usually interpret this pointer setting as indicating that the value of the measured dynamical variable is $B_j$, but, as we will see in the next section, this inference is not licenced under the Q-based interpretation.) The experimentalist will ascribe a probability distribution
\begin{equation}
Q_j(\boldsymbol\alpha)\,
\end{equation}
whose support is (almost entirely) confined to $\Gamma_j$. To show that this function $Q_j$ is (or can be approximated well by) the Q-function of some density operator $\hat\rho_j$, note that the Q-functions associated with coherent states $|\boldsymbol{\alpha'}\rangle$ corresponding to phase space points $\boldsymbol{\alpha'}$ are Gaussians centred around $\boldsymbol{\alpha'}$:
\begin{eqnarray}
Q_{\boldsymbol{\alpha'}}(\boldsymbol\alpha)&=&\frac{1}{\pi^N}\exp(-|\boldsymbol\alpha-\boldsymbol{\alpha'}|^2)\,.\label{coherentstateQ}
\end{eqnarray}
Now recall that $\Gamma_j$ is macroscopic, whereas the Gaussians $Q_{\boldsymbol{\alpha'}}$ in Eq.\ (\ref{coherentstateQ}) fall off on the scale of $\hbar$. This makes it reasonable to assume that $Q_j$ can be approximated to a high degree of precision by a suitably weighted integral of the $Q_{\boldsymbol{\alpha'}}$ associated with phase space points $\boldsymbol{\alpha'}\in\Gamma_j$, weighted by some function $\mu_j(\boldsymbol{\alpha'})$ which, as well, is non-zero only in $\Gamma_j$:
\begin{eqnarray}
Q_j(\boldsymbol\alpha)&\approx&\int_{\boldsymbol{\alpha'}\in\Gamma_j}\mu_j(\boldsymbol{\alpha'})Q_{\boldsymbol{\alpha'}}(\boldsymbol\alpha)d\boldsymbol{\alpha'}\nonumber\\
&=&\frac{1}{\pi^N}\int_{\boldsymbol{\alpha'}\in\Gamma_j}\mu_j(\boldsymbol{\alpha'})\langle\boldsymbol\alpha|\boldsymbol{\alpha'}\rangle\langle\boldsymbol\alpha'|\boldsymbol\alpha\rangle d\boldsymbol{\alpha'}\\
&=&\frac{1}{\pi^N}\langle\boldsymbol\alpha|\left(\int_{\boldsymbol{\alpha'}\in\Gamma_j}\mu_j(\boldsymbol{\alpha'})|\boldsymbol{\alpha'}\rangle\langle\boldsymbol{\alpha'}| d\boldsymbol{\alpha'}\right)\boldsymbol\alpha\rangle.\nonumber\\
&=&\frac{1}{\pi^N}\langle\boldsymbol\alpha|\hat\rho_j|\boldsymbol\alpha\rangle.\nonumber
\end{eqnarray}
That is, we can approximate $Q_j$ by the Q-function associated with the density operator
\begin{equation}
\hat\rho_j=\int_{\boldsymbol\alpha\in\Gamma_j}\mu_j(\boldsymbol\alpha)|\boldsymbol\alpha\rangle\langle\boldsymbol\alpha| d\boldsymbol\alpha\,.\label{rho_j}
\end{equation}
This means that, if $S$ is prepared such that the quantum state assigned to it is $|B_j\rangle$, the post-measurement quantum state assigned to the apparatus $A$ must be such a density operator $\hat\rho_j$. Accordingly, if we denote the pre-measurement quantum state of $A$ by $\hat\rho_0$, the measurement interaction must fulfil
\begin{equation}
|B_j\rangle\langle B_j|\hat\rho_0\mapsto|B_j\rangle\langle B_j|\hat\rho_j\,.\label{premeasure}
\end{equation}
Now, in the general case the pre-measure\-ment state of $S$ is some superposition
\begin{eqnarray}
\sum_ic_i|B_i\rangle,
\end{eqnarray}
and the $S+A$ pre-measurement quantum state is
 \begin{eqnarray}
\sum_{i,j}c_j^*c_i|B_i\rangle\langle B_j|\hat\rho_0,\,.
\end{eqnarray}
The measurement apparatus $A$ is supposed to be macroscopic and it has a mascroscopic environment. This means that environment-induced decoherence must be taken into account in the measurement interaction. Due to decoherence, the post-measurement (reduced) density operator of the combined system $S$+$A$ will be approximately
 \begin{eqnarray}
\hat\rho_{S+A}=\sum_{i}|c_i|^2|B_i\rangle\langle B_i|\hat\rho_i\,.\label{postmeasure}
\end{eqnarray}
Let us refer to the Q-function of this $\hat\rho_{S+A}$ as $Q_{S+A,post}$.

The post-measurement Q-function $Q_{A,post}$ of the apparatus $A$ is obtained by integrating this $Q_{S+A,post}$ over the phase space of $S$ or, alternatively, from the post-measurement reduced density operator of $A$
 \begin{eqnarray}
\hat\rho_{red,A}=\sum_i |c_i|^2\hat\rho_i\,,
\end{eqnarray}
which, for the Q-function of the apparatus at the end of measurement, yields
 \begin{eqnarray}
Q_{post,A}(\boldsymbol\alpha)&=&\frac{1}{\pi^N}\langle\boldsymbol\alpha|\left(\sum_i |c_i|^2\hat\rho_i\right)|\boldsymbol\alpha\rangle\\
&=&\sum_i |c_i|^2\frac{1}{\pi^N}\langle\boldsymbol\alpha|\hat\rho_i|\boldsymbol\alpha\rangle\\
&=&\sum_i |c_i|^2Q_i(\boldsymbol\alpha)\,.
\end{eqnarray}
Now, according to the Q-based interpretation, the probability $P(j)$ of obtaining a `pointer setting' in $\Gamma_j$ is given by a phase space integral of this Q-function over $\Gamma_j$. That integral overwhelmingly comes from the term containing $Q_j$, and the integral of $Q_j$ over $\Gamma_j$ is very nearly $1$, so the result is, as announced in Eq.\ (\ref{announce}):
 \begin{eqnarray}
P(j)&=&\int_{\boldsymbol\alpha\in\Gamma_j} d\boldsymbol\alpha\ Q_{red,A}(\boldsymbol\alpha)\nonumber\\
&\approx&|c_j|^2.
\end{eqnarray}
This is the familiar result as well-confirmed for quantum mechanics using Weyl quantization. It has been recovered here based on reasonable-looking assumptions about the measurement process---notably, that the $\Gamma_i$ are macroscopic---and on the interpretation of the Q-function as a proper probability distribution. So we seem to have some reason to believe that the Q-based interpretation may indeed be empirically adequate. \citet{drummondreid} discuss some further special cases, namely, measurement of a quadrature of the electromagnetic field through amplification, measurement of single-particle spin, and measurement of EPRB-type correlations using spins of entangled particles. In their analyses, the measurement apparatus plays the role of an amplification device, and decoherence is not invoked beyond amplification. Drummond and Reid come to the conclusion that interpreting the Q-function as a proper probability function yields empirically adequate consequences in all the settings that they consider.

\section{A Further Look at Measurement}

One may have a further worry about the coherence of the Q-based interpretation:

The Q-functions of any two orthogonal quantum states $|\psi_1\rangle$ and $|\psi_2\rangle$ in general have overlapping support in phase space. For example, for the one-dimensional harmonic oscillator with complex phase space variable $\beta$ (in analogy to $\alpha$ in Section 2 -- but I will reserve `$\alpha$' for the phase space variable of the apparatus) the Q-functions associated with eigenstates $|n\rangle$ (with $n=1, 2, ...$) of the Hamiltonian are
\begin{eqnarray}
Q_{|n\rangle}(\beta)=\frac{|\beta|^{2n}}{n!}\exp\left(-|\beta|^2\right)\,.
\end{eqnarray}
Especially for values of $n$ that are close to each other, any two of these functions can have considerable `overlap', i.e. there are regions in phase space where they are both (significantly) different from zero.

Now assume that a quantum system is prepared in an eigenstate $|B_1\rangle$ of some dynamical variable $\mathcal B$, associated with an operator $\hat B$, and that, in line with the Q-based interpretation, this preparation results in the system being at some phase space point $\beta_1$ where $Q_{|B_1\rangle}$ is non-zero. Now, if the system does not undergo any further interactions after measurement, subsequent measurement of $\mathcal B$ results, with probability $1$, in the outcome being the eigenvalue $B_1$ for which $\hat B|B_1\rangle=B_1|B_1\rangle$. In other words, the probability of obtaining $B_1$ when measuring $\mathcal B$ and the system is at $\beta_1$ is $1$:
\begin{eqnarray}
P(B_1|\beta_1,\mathcal B)=1\,. \label{contradiction1}
\end{eqnarray}
But this cannot be right. Inevitably, because the Q-functions of orthogonal states can have non-trivial overlap, there will also be a different eigenstate $|B_2\rangle$ of $\hat B$ such that also $Q_{|B_2\rangle}(\beta)$ is non-zero at $\beta_1$. But this means that $\beta_1$ could as well have resulted from preparing $|B_2\rangle$. And, in that case, as we know from the empirical success of quantum mechanics, when measuring $\mathcal B$ we could be sure to obtain the result $B_2$. Thus, by the same reasoning that leads to Eq.\ (\ref{contradiction1}) we derive
\begin{eqnarray}
P(B_2|\beta_1,\mathcal B)=1\,. \label{contradiction2}
\end{eqnarray}
Since, by assumption, $B_1\neq B_2$, Eqs.\ (\ref{contradiction1}) and (\ref{contradiction2}) are in contradiction with each other.

This should not make the proponent of the Q-based interpretation nervous. Her reaction should simply be to reject both Eq.\ (\ref{contradiction1}) and Eq.\ (\ref{contradiction2}). The correct calculation to determine the probabilities of the different measurement outcomes is different. It uses the Q-function of the combined system $S+A$, which, after taking into account the measurement interaction but before taking into account the measured result, is (see Eqs.\ (\ref{postmeasure}) and (\ref{rho_j}))
\begin{eqnarray}
\hat\rho_{S+A}&=&\sum_{i}|c_i|^2|B_i\rangle\langle B_i|\hat\rho_i\nonumber\\
&=&\sum_{i}|c_i|^2|B_i\rangle\langle B_i|\int_{\boldsymbol\alpha\in\Gamma_i}\mu_i(\boldsymbol\alpha)|\boldsymbol\alpha\rangle\langle\boldsymbol\alpha| d\boldsymbol\alpha\,.\label{diagonalrho}
\end{eqnarray}
To this density operator corresponds the Q-function
\begin{eqnarray}
Q_{S+A}(\beta,\boldsymbol\alpha)&=&\frac{1}{\pi^{N+1}}\sum_i|c_i|^2|\langle\beta|B_i\rangle|^2|\int_{\boldsymbol{\alpha'}\in\Gamma_i} |\mu_i(\boldsymbol{\alpha'})|^2|\langle\boldsymbol \alpha|\boldsymbol{\alpha'}\rangle|^2 d\boldsymbol{\alpha'}\nonumber\\
&=&\sum_i|c_i|^2Q_{|B_i\rangle}(\beta)\ \int_{\boldsymbol{\alpha'}\in\Gamma_i} |\mu_i(\boldsymbol{\alpha'})|^2Q_{|\boldsymbol{\alpha}\rangle}(\boldsymbol\alpha') d\boldsymbol{\alpha'}\,. \label{outcomeprobs}
\end{eqnarray}
In the Q-based interpretation, this is interpreted as a proper probability distribution over $(\beta,\boldsymbol\alpha)$, assigned before registering the pointer position. Registering the pointer position means finding the phase space location of $A$ to be within some macroscopic phase space region $\Gamma_j$. Performing Bayesian updating on this information---that the value of $\boldsymbol\alpha$ lies in $\Gamma_j$---means switching from assigning the Q-function Eq.\ (\ref{outcomeprobs}) to assigning
\begin{eqnarray}
Q_{S+A}^{j}(\beta,\boldsymbol\alpha)&=&\mathcal N\ Q_{|B_j\rangle}(\beta)\ \int_{\boldsymbol{\alpha'}\in\Gamma_j} |\mu_j(\boldsymbol{\alpha'})|^2Q_{|\boldsymbol{\alpha}\rangle}(\boldsymbol{\alpha'}) d\boldsymbol{\alpha'}\,.\label{Qupdate}
\end{eqnarray}
Here $\mathcal N$ is a normalization factor chosen such that the integral of $Q_{S+A}^{j}(\beta,\boldsymbol\alpha)$ over the combined system $S+A$ phase space is $1$. The switch to Eq.\ (\ref{Qupdate}) entails measurement collapse for the quantum state of the measured system $S$. It gets updated from $\sum_ic_i|B_i\rangle$ to $|B_j\rangle$. Note that this is a nice result: von Neumann collapse for the measured system state has been derived as a consequence of applying ordinary Bayesian updating to the apparatus Q-function when conditioning on the registered pointer location.

The eigenvalue $B_j$ of the eigenstate $|B_j\rangle$ to the operator $\hat B$ has not appeared at all in this analysis. It is only indirectly `inferred' by the experimentalist as the pointer location is found to be in $\Gamma_j$ and the measured system state is `collapsed' to $|B_j\rangle$. But even talking of $B_j$ as `inferred' might be misleading because, in general, for values of $\beta$ where $Q_{|B_j\rangle}(\beta)$ is non-zero, the true value of $\mathcal B$ will not be $B_j$. In the Q-based interpretation, it is simply not the case that the possible values of $\mathcal B$ are the eigenvalues of $\hat B$. This is analogous to de Broglie-Bohm theory. As in de Broglie-Bohm theory, the expression `measurement of dynamical variable $\mathcal B$' must be taken with a grain of salt and should not be understood as `revealing the true value of $\mathcal B$'.

Let us come back to the scenario that gave rise to the worry discussed in the beginning of this section. There, a measured system was assumed to be prepared in $|B_1\rangle$ and then undergoing measurement of the dynamical variable $\mathcal B$. In that situation, the sum over $i$ in Eq.\ (\ref{outcomeprobs}) becomes trivial and the update to Eq.\ (\ref{Qupdate}) does not change the Q-function of $S+A$. The Q-function of $S$ also stays the same, namely $Q_{|B_1\rangle}(\beta)$, throughout the measurement interaction and after registering the pointer location.

Thus, the correct take, according to the Q-based interpretation, is not that the value of $\mathcal B$ is $B_1$ both before and after the measurement---depending on the value of $\beta$, the actual value of $\mathcal B$ may actually be different at any of those times---but that the Q-function assigned to the measured system does not change when that system is subjected to (what we call) measurement of the dynamical variable $\mathcal B$ with respect to which $S$ is prepared in an eigenstate $|B_1\rangle$.

The answer to the worry with which this section started, thus, is that, in order to make claims about a quantum system's Q-function at different times $t$, the Q-function evolution itself must be considered in full. Effects from interactions with other systems, be they `measurement' interactions or others, must be included in terms of how they affect the time-evolution of Q.

We can now see more clearly what went wrong in the derivation of the jointly inconsistent Eqs.\ (\ref{contradiction1}) and (\ref{contradiction2}): if the measured system is prepared in some state $|C\rangle$---e.g. by `measuring' a dynamical variable $\mathcal C$ and selecting for the outcome $C$---then this preparation history affects the Q-function and, therefore, has to be taken into account when computing the probabilities of the different possible inferred values $B_1$, $B_2$, ... Those are not screened off from the preparation procedure by the phase space location $\beta$ that happens to result from the preparation procedure. It is not acceptable to assume that $P(B_j|\beta,\mathcal B,C,\mathcal C)$ can be simplified according to
\begin{eqnarray}
P(B_j|\beta,\mathcal B,C,\mathcal C)= P(B|\beta,\mathcal B)\,,\label{lambdamediation1}
\end{eqnarray}
and the derivation of Eqs.\ (\ref{contradiction1}) and (\ref{contradiction2}) is blocked.

The following section discusses what rejecting (\ref{lambdamediation1}) means and whether it should be considered acceptable.

\section{$\lambda$-mediation and Its Possible Violation}

Perhaps the main motivation for the Q-based interpretation is that it allows one to interpret quantum mechanics along realist lines, somewhat akin to classical statistical mechanics. In classical statistical mechanics one usually does not interpret the phase space probability distribution $\rho$ as an objective property of the system. So, if one wants the parallel between quantum mechanics according to the Q-based interpretation and quantum statistical mechanics to be far-reaching, one will interpret the Q-function as `non-ontic' and the system's phase space location $\beta$ as its complete `ontic' physical state. This is also the perspective that \citet{drummondreid}, pioneers of the idea of interpreting the Q-function as a proper probability function, have in mind, as will be discussed in the following section. The wave function uniquely fixes the Q-function and vice versa, so, plausibly an ontic (non-ontic) view of the Q-function is equivalent to an ontic (non-ontic) view of the quantum state. At the end of this section I consider the option of combining the Q-based interpretation with an ontic view of $Q$, but let us first consider the consequences of a non-ontic view of $Q$.

In that case, accepting that Eq.\ (\ref{lambdamediation1}) fails amounts to rejecting what \citet{leiferpusey} refer to as `$\lambda$-mediation': the assumption that a system's ontic state $\lambda$ `screens off' the outcomes of measurements of the system from its preparation history. Following \citet[p.\ 10]{leiferpusey}, $\lambda$-mediation is the statement:
\begin{quote}
The ontic state $\lambda$ mediates any remaining correlation between the preparation and the measurement[.]
\end{quote}
In terms of the symbols used in this paper, the formula that according to Leifer and Pusey expresses $\lambda$-mediation is precisely Eq.\ (\ref{lambdamediation1}), replacing $\beta$ with the general ontic state variable $\lambda$:
\begin{eqnarray}
P(B|\lambda,\mathcal B,C,\mathcal C)=P(B|\lambda,\mathcal B)\,.\label{lambdamediation}
\end{eqnarray}
Is giving up $\lambda$-mediation an unacceptably high price for embracing the Q-based interpretation?

I don't think so. In fact, a recent paper by \citet{adlam} makes a strong independent case against requiring $\lambda$-mediation. To understand Adlam's point, it is important to appreciate that $\lambda$-mediation can be construed as an assumption of temporal locality: if Eq.\ (\ref{lambdamediation}) holds, any correlation, and a fortiori any causal influence, between the preparation setting $\mathcal C$ and outcome $C$ on the one hand and the measurement outcome $B$ must be mediated by the temporally intermediate ontic state $\lambda$.

However, as Adlam points out, we already know quantum theory to be incompatible with spatial locality: Eq.\ (\ref{lambdamediation}), but interpreted such that the setting $\mathcal C$ and outcome $C$ are space-like separated from $\mathcal B$ and $B$. This follows from the fact that quantum correlations violate Bell inequalities. (These correlations are recovered in the Q-based interpretation. See the next section for some more comments.) And, as Adlam sees it, there are no compelling reasons for insisting on temporal locality while accepting spatial non-locality. In fact, as shown by \citet{evansetal}, accepting temporal non-locality is unavoidable if one is neither prepared to accept retrocausality nor that operational and ontic symmetries may come apart.

Adlam notes that, as will be discussed below, temporal locality in the form of $\lambda$-mediation is used in the derivation of `$\psi$-ontology' theorems, which entail that quantum states must be states of reality. She regards this conclusion as unattractive and concludes that pursuing temporally non-local interpretations is potentially the main neglected, yet promising, route in the foundations of quantum theory:
\begin{quote}
[F]ully embracing temporal nonlocality might open up untapped possibilities for the interpretation of quantum theory, and hence the whole landscape of quantum foundations becomes markedly different when temporal nonlocality is taken seriously. (\citet{adlam}, p.\ 3)
\end{quote}
To be sure, the assumption that ontic states mediate between preparation and measurement may indeed seem very natural. Notably, Leifer and Pusey themselves regard $\lambda$-mediation as encoding a core feature of scientific realism:
\begin{quote}
	In our view, the idea that ontic states are responsible for correlations between preparation and measurement, which is the idea behind $\lambda$-mediation, is also a core feature of a realist theory. It encodes the idea that ontic states are supposed to explain what we see in experiments. (\citet{leiferpusey}, p.\ 21)
\end{quote}
And indeed, explaining what we see in experiments in terms of the ontic states of microsystems is part and parcel of physical practice. But this is not a strong reason for accepting $\lambda$-mediation: if our best fundamental physical theory entails strong nomic constraints on correlations across time but rules out that those correlations are mediated by temporally intermediate ontic states, this indicates that fundamental physics is interestingly different from everyday physics and the higher-level sciences; it does not indicate that we have to scale back the ambitions of science. 

When adopting the Q-based interpretation, one can avoid giving up $\lambda$-mediation by adopting an ontic account of $Q$ and denying that a system's phase location is its complete ontic state. The complete ontic state would then be $\Lambda=(\beta_0,Q(\beta))$ (or, equivalently, $\Lambda=(\beta_0,\hat\rho)$). Borrowing terminology from certain modal interpretations of quantum mechanics, one might refer to a system's phase space location $\beta_0$ as its value state and its Q-function $Q(\beta)$ as its dynamical state: the latter determines the probabilities of specific values of $\beta$ at different times, and these cannot be determined from the value state $\beta_0$ at some specific time $t$ alone.

However, alongside the advantage of allowing one to preserve $\lambda$-mediation, an ontic view of the Q-function also has the severe downside that it creates a schism between probabilities as actually used in the application of quantum mechanics and the Q-function:\footnote{In de Broglie-Bohm theory, this schism manifests itself in the challenge of justifying why `quantum equilibrium' obtains. In the Everett interpretation, it manifests itself in the probability problem---justifying why branch weights effectively manifest themselves as probabilities---which is often regarded as the most serious difficulty for the Everett interpretation.} whenever some measurement outcome is registered, the Q-function assigned to the measured system gets `updated', in the setting considered above from $Q_{|\psi\rangle}(\beta)$ to $Q_{|B_i\rangle}(\beta)$. A proponent of the Q-based interpretation who regards the Q-function as ontic will presumably adopt a `no- (ontic) collapse' view and regard this update as not corresponding to any physical change. Like the proponent of de Broglie-Bohm theory, she will consider measurement collapse as a shift to an `effective' Q-function, made for the sake of predictive and computational convenience, and identify the proper, physical, Q-function with the one corresponding to the uncollapsed quantum state. Strictly speaking, the only quantum state that is really ontic, in this view, is the uncollapsed `ur-' wave function of the universe that includes all decoherent branches, including `empty' ones that have no further effect on the dynamics of phase space location.\footnote{However, if experiments that involve the reversal of decoherence (`recoherence') ever become a reality, calculations based on the collapsed state and its associated Q-function may become empirically inadequate. Based on this observation, \citet{lazarovici} have recently argued that the (hypothetical) possibility of recoherence counts decisively against epistemic accounts of the quantum state. Recoherence also plays a central role in the much-discussed paradox recently proposed by \citet{frauchigerrenner}. The response of the proponent of the Q-based interpretation to that paradox will plausibly depend on the ontological status accorded to $Q$.}

One may also feel that it is unclear what kind of entity the Q-function would have to be on an ontic view of it. Options here seem to largely parallel those for interpreting the non-collapsing universal wave function $\psi$ in de Broglie-Bohm theory, which unavoidably has `empty branches'.

In any case, it appears to be more straightforward that the global $\psi$ in de Broglie-Bohm theory must be ontic than that the uncollapsed, global, Q-function must be ontic in the Q-based interpretation. The reason is that in de Broglie-Bohm theory, $\psi$ directly couples to ontic particle velocity via the guidance equation for the position coordinates $\boldsymbol Q_k$:
\begin{eqnarray}
\frac{d\boldsymbol Q_k}{dt}=\frac{\hbar}{m_k}\Im \frac{\psi^*\partial_k\psi}{\psi^*\psi}(\boldsymbol Q_1, ..., \boldsymbol Q_N)\,.
\end{eqnarray}
Inasmuch as Eq.\ (\ref{path}) is a good guide to microdynamics in the Q-based interpretation, the situation is different there: the `time-mixing' conditional probabilities of field configurations at different times do not involve the Q-function but only the central difference Lagrangian $\mathcal L$. There does not seem to be a similarly straightforward reason for adopting a $\psi$-ontic (Q-ontic) view as for de Broglie-Bohm theory. However, assuming that some specific Q-function has been assigned for some specific time and assuming that no further updating takes place, it is an objective matter which Q-function one should assign for any other specific time. In that sense, the central difference Langrangian $\mathcal L$ and the stochastic diffusion equations governing the time-evolution of the Q-function are plausibly objective in any version of the Q-based interpretation, including Q-epistemic ones. (Whether one may want to call them `ontic' will presumably depend on one's preferred metaphysical account of laws of time-evolution.)

\section{Evading No-go Theorems and the Question of Retrocausality}

Epistemic accounts of quantum states (`$\psi$-epistemic accounts') are prima facie attractive because they allow one to conceive of measurement collapse as related to Bayesian updating (\citet{spekkens_epistemic,fuchsschack,friederich,healey}). The Q-based interpretation, when combined with an epistemic account of the Q-function, exemplifies this attractive feature. It is expressed in the update of Eq.\ (\ref{outcomeprobs}) to Eq.\ (\ref{Qupdate}). However, there are various no-go theorems (\citet{pbr,hardy,colbeckrenner}) which seriously limit the options for viable $\psi$-epistemic accounts. One may wonder whether they pose any problems for the Q-based interpretation.

Interestingly, as already mentioned, $\lambda$-mediation is used in the proofs of all those theorems. The streamlined review (\citet{leifer}) of the `$\psi$-ontology' theorems uses the ontological models framework (\citet{harriganspekkens}) as the unifying background. But this framework includes $\lambda$-mediation as a conceptual presupposition of ontological models. Theorems proved in the ontological models framework are thus inapplicable to the Q-based interpretation. Those theorems, including the famous PBR theorem, therefore do not spell trouble for the Q-based interpretation when combined with an epistemic account of $Q$.

$\lambda$-mediation is also used in proofs of Bell's theorem. It appears in Bell's original proof (\citet{belltheorem}, Eq.\ (1)) and implicitly in modern versions as a pre-condition for locality, for instance (\citet{gntz}, Eq.\ (4)). Thus, trivially, Bell locality is violated by the Q-based interpretation. There is no particular difficulty with recovering the observed violations of Bell inequalities in the Q-based interpretation, as demonstrated by \citet[Fig.\ 5]{drummondreid} for a simple special case.

The fact that $\lambda$-mediation is violated in the Q-based interpretation when combined with an epistemic account of $Q$ also entails that the standard characerization of (non-) contextuality due to \citet{spekkens} is not applicable to this interpretation. The same holds for a more recent characterization of non-contextuality (\citet{shrapnelcosta}), which is also based on a version of $\lambda$-mediation (`$\omega$-mediation'). Inasmuch as one regards any interpretation that violates $\lambda$-mediation (or $\omega$-mediation) as trivially contextual, one will regard the Q-based interpretation combined with an epistemic account of $Q$ as trivially contextual. 

The Q-based interpretation combined with an ontic view of $Q$ does not violate $\lambda$-mediation and can be classified according to the characterization of \citet{spekkens}. Orthodox quantum mechanics with only pure states regarded as ontic (the `Beltrametti-Bugajski model') qualifies as preparation contextual by that standard because different combinations of pure state preparations can result in different distributions over ontic states even if they correspond to the same mixed state $\hat\rho$. The same holds for de Broglie-Bohm theory and, plausibly, for the Q-based interpretation when combined with an ontic account of $Q$. Both these interpretations have analogous conceptions of a global $\Psi$/$Q$ and of collapse as merely effective and approximate.

\citet{drummondreid}, proposing the idea of interpreting the Q-function as a proper probability function, clearly aim to combine it with an epistemic account of quantum states when they write that `that the originators of quantum mechanics regarded the quantum wave-function as statistical [50] for good reason'. (\citet{drummondreid}, p.\ 3) Unlike the present paper, however, they do not consider $\lambda$-mediation and its potential failure to be relevant as enabling a $\psi$-epistemic account. In the first published preprint version of (\citet{drummondreid}) they argue that the reason why the PBR and other `$\psi$-ontology' theorems do not apply to their interpretation, according to themselves, is that this interpretation is retrocausal: `Theorems requiring an ontological wave-function do not apply to our model [...] because of retrocausality [...] due to negative diffusion terms in the dynamical equations ' (\citet{drummondreid} v1, p.\ 1). This is an allusion to Q-function dynamics in bosonic quantum field theory being determined by diffusion equations with opposite signs for different degrees of freedom, as discussed in the context of Eqs.\ (\ref{diffusion}) and (\ref{path}).

Drummond and Reid are correct that retrocausality, just like the failure of $\lambda$-mediation, can make $\psi$-ontology theorems inapplicable. (See (\citet{friederichevans}) for a recent review of the options opening up in quantum foundations when accepting retrocausality.) Notably, retrocausal effects from measurement settings backwards in time to earlier states of affairs can lead to violations of an assumption called measurement independence. This assumption is incorporated in the ontological models framework just like $\lambda$-mediation and was identified as crucial already by Bell in the derivation of his famous non-locality theorem. Measurement independence obtains if measurement settings are uncorrelated with earlier states of affairs $\lambda$ that are considered as possible confounders. Giving up measurement independence and accepting retrocausality might appear to be an extravagant move, but there are strong arguments that it is actually metaphysically `cheap' (\citet{evans}).

However, the appearance of negative diffusion terms in Q-function dynamics does not by itself entail a failure of measurement independence. These are logically independent matters. If a system is prepared at time $t_0$ using a measurement setting $\mathcal C$ and measurement output $C$ and is subjected to measurement at time $t_f>t_0$ using a measurement setting $\mathcal B$, measurement independence comes down to the identity
\begin{eqnarray}
P_t(\lambda|\mathcal B,C,\mathcal C)=P_t(\lambda|C,\mathcal C)\label{MI}
\end{eqnarray}
for intermediate times $t$ with $t_0<t<t_f$. According to Eq.\ (\ref{MI}), conditioning on a later measurement setting has no effect on the probability assigned to the ontic state $\lambda$ at $t$ when controlling for the preparation procedure.

Eq.\ (\ref{MI}) holds in de Broglie-Bohm theory, where the ontic state $\lambda$ includes both the position of the system in configuration space and the universal quantum state $\Psi$ (which evolves deterministically anyway). There is no reason to think that the situation is any different in the no-collapse $\psi$-ontic (or Q-ontic) version of the Q-based interpretation, where $\lambda$ includes the phase space location $\beta$ and the ontic $Q$.

In an epistemic account of $Q$, the time-evolution of $Q$ does not occur on the ontic level of physical processes (though the laws of time-evolution themselves are arguably objective, as argued at end of Section 6), so the fact that it is partly governed by negative diffusion coefficients provides even less reason for inferring a violation of measurement independence than it does in an ontic account of $Q$. This still leaves open the possibility that there could be an argument for measurement independence being violated in epistemic accounts of $Q$, but I do not currently see one.\footnote{I would like to thank an anonymous reviewer for curing me of the impression to possess such an argument.}

Depending on one's preferred philosophical account of causality, one may diagnose retrocausality without violation of measurement independence. In principle, there could be accounts of causality according to which the fact that Q-function dynamics is characterized by backward-in-time diffusion for some degrees of freedom intrinsically qualifies as an instance of retrocausality. But then again, retrocausality identified by the standards of such accounts does not entail the violation of measurement independence and, therefore, is not directly linked to the `retrocausality loophole' of the no-go theorems.

On the other hand, retrocausality by the standards of mainstream interventionist accounts of causality does seem to require the violation of measurement independence, i.e. that Eq.\ (\ref{MI}) does not hold and that choosing which $\mathcal B$ to measure has an impact on $\lambda$. The fact that we have a formula Eq.\ (\ref{path}) to compute the `time-mixing' conditional probability $P(\phi_{x,f},\phi_{y,0}|\phi_{x,0},\phi_{y,f})$, which encodes the stochastic dynamics of diffusion, does not mean that we can in practice manipulate $\phi_{y,f}$ and thereby, indirectly, intervene backwards in time on $\phi_{y,0}$. Put differently, the fact that, as \citet[p.\ 7]{drummondreid} highlight, future boundary constraints are needed `to obtain a single, probabilistic, trajectory', does not mean that agents can indeed, as a matter of practical possibility, impose future conditions and thereby affect the past. Without supplementary arguments to the contrary, Eq.\ (\ref{path}), as it stands, has a purely correlational interpretation.

To conclude, it might be possible to provide a compelling argument that the Q-based interpretation entails retrocausality, as claimed by Drummond and Reid, but so far no such argument has been given.

\section{Towards Q-based Interpretations of Relativistic Quantum Theories and Quantum Field Theories}
For historical reasons---and for the sake of simplicity---interpretations of quantum theories are often introduced for non-relatistic quantum mechanics. Except for the passages where I reviewed Drummond's results centred around Eqs.\ (\ref{diffusion}) and (\ref{path}) I have done the same here for the Q-based interpretation. For some interpretations, notably the de Broglie-Bohm theory, the main technical challenge that they face is precisely how to generalize them to relativistic and/or quantum field theoretical settings.

Fortunately, the prospects for applying the Q-based interpretation to relativistic quantum field theories seem quite good. In fact, \citet{drummondreid} introduce the idea of interpreting the Q-function as a proper probability for relativistic quantum field theory, with an ontology of classical fields rather than point-like particles. In that context, the Q-function $Q(\lambda)$ is defined over composite classical fields
\begin{eqnarray}
\lambda=[\phi_1, \phi_2, ...,\xi_1, \xi_2, ...]\,,
\end{eqnarray}
where $\phi_1$, $\phi_2$, ... correspond to bosons, and  $\xi_1$, $\xi_2$, ... are real, anti-symmetric, matrix-valued fields which correspond to fermions (\citet{drummondreid}, p.\ 2).

As Drummond and Reid highlight, `[t]he Q-function is the probability at some time $t$, but in
this model, fields have continuous trajectories $\lambda(t)$ defined at all times' (\citet{drummondreid}, p.\ 3). Because $Q$ assigns a value to any given field configuration and its conjugate momentum at a given time, $Q$ itself is frame-dependent. On an epistemic account of quantum states---which, as discussed above, may fit most naturally with the Q-based interpretation---different agents may legitimately assign different quantum states to one and the same system. If those agents are sufficiently far apart and/or their rest frames are different, the information that they have about the system, on which they can base their quantum state assignment, may be quite different. This observation raises intriguing questions about the constraints under which Q-function ascriptions by different agents to fields in one and the same space-time region can be coherent and legitimate. Addressing those questions is, however, beyond the scope of the present exploration of the Q-based interpretation.

In any case---and potentially relevant to the question of how the Q-based interpretation fits into a relativistic setting---the Q-based interpretation does not seem to force one to accept any superluminal causation, at least not absent any further insights into micro-physics that would suggest otherwise. Notably, if the operators corresponding to dynamical variables associated with space-like separated regions commute (`local commutativity'), the reduced density operator $\hat\rho_{red}$ of a system confined to some (finite, compact) region $I$ is not affected by the interactions involved in measuring some dynamical variable $\mathcal B$ in some region $II$ that is space-like separated from $I$.

For assume that a system in region $I$ has been prepared in accordance with a procedure associated with some quantum state $\hat\rho$. Then, according to the Q-based interpretation, the probabilities of its possible phase space locations $\beta$, assigned from the perspective of an agent co-located in $I$, are
\begin{eqnarray}
P_I(\beta)=Q_{\hat\rho}(\beta)\,.
\end{eqnarray}
Since, per local commutativity, the measurement of $\mathcal B$ in region $II$ has no instantaneous effect on $\hat\rho$, conditioning with respect to its setting has no effect on the probabilities in region $I$:
\begin{eqnarray}
P_I(\beta|\mathcal B)=Q_{\hat\rho_{red}}(\beta)=Q_{\hat\rho}(\beta)=P_I(\beta)\,.
\end{eqnarray}
Since the measurement setting $\mathcal B$ can be seen as an arbitrary intervention set to some variable, this indicates that interventions cannot influence phase space locations of objects at space-like distance according to the Q-based interpretation. Of course, conditioning not only on the measurement setting $\mathcal B$ but also on the measurement outcome $B$ can have a dramatic effect on $\hat\rho_{red}$ and, thus, on the probabilities of phase space locations of objects at space-like distance. But these outcomes are not manipulable and, hence, the `non-local' correlations violating Bell-type inequalities are not causal, at least not by the standards of mainstream interventionist accounts of causation.

A further reassuring finding when it comes to generalizing the Q-based interpretation to relativistic quantum field theory is that there seem to be unique Q-functions in both bosonic and fermionic quantum field theories, including those with massive fields. \citet{Rosales-Zarate} provide an abstract characterization of a Q-function in terms of fulfilling the following three criteria:
\begin{quote}
(1) It exists uniquely for any quantum density-matrix.

(2) It is a positive probability distribution.

(3) Observables are moments of the distribution.\\(\citet{Rosales-Zarate}, p.\ 2)
\end{quote}
The general form of a Q-function that fulfils these criteria is (\citet{Rosales-Zarate}, Eq.\ (2.2))
\begin{eqnarray}
Q(\boldsymbol\lambda)=Tr\large\lbrack\hat\Lambda(\boldsymbol\lambda)\hat\rho\large\rbrack\,,
\end{eqnarray}
where $\Lambda(\boldsymbol\lambda)$ is a positive definite Hermitian Hilbert space basis which is normalized such that, for an integral over phase space domain $\mathcal D$ with phase space measure $d\mu(\boldsymbol\lambda)$ (\citet{Rosales-Zarate}, Eq.\ (2.1)),
\begin{eqnarray}
\int_\mathcal D \hat\Lambda(\boldsymbol\lambda)d\mu(\boldsymbol\lambda)=\boldsymbol 1\,.
\end{eqnarray}
For bosonic field theories, the basis $\Lambda(\boldsymbol\lambda)$ for which the three above conditions are uniquely fulfilled consists of projectors onto the Glauber coherent states which generalize the harmonic oscillator coherent states Eq.\ (\ref{coherentstate}) to coherent states of the boson field. For fermionic field theories, fermionic Gaussian operators take the place of the projectors onto the coherent states (\citet{Rosales-Zarate}, Eqs.\ (3.6), (3.7)), the Q-function basis has a `gauge' freedom, but with these replacements the overall form of the Q-function is same as for boson fields (\citet{Rosales-Zarate}, Eqs.\ (4.18), (5.1)).

\section{Conclusion and Outlook at Further Challenges}
I conclude this paper by recapitulating selected features of the Q-based interpretation that make it attractive and worthy of being developed further. I also outline challenges for future work.

First, an incomplete list of attractive features of the Q-based interpretation:
\begin{itemize}
\item The Q-based interpretation promises having no measurement problem. It assumes that every quantum system has a definite location in phase space at all times and all dynamical variables defined as functions on phase space have definite values at all times.
\item The Q-based interpretation treats microscopic and macroscopic systems as on the same ontological footing and does not rely on any primitive anthropocentric concepts (such as `measurement' or `observer').
\item The Q-based interpretation allows one to interpret wave function collapse as reflecting Bayesian updating of the probability distribution over phase space in the light of evidence about the pointer location (though collapse is understood as merely `effective' if an ontic account of $Q$ is chosen).
\item The Q-based interpretation does not add any novel elements to the formalism of quantum theory that are not needed to derive that formalism's empirical consequences. Notably, it does not postulate any additional dynamical principles of wave function collapse. The micro-physics encoded in Eq.\ (\ref{path}) follow from Q-function dynamics, which in turn follow from the von Neumann equation.
\item The Q-based interpretation applies beyond non-relativistic quantum mechanics, to relativistic and field-theoretic settings. It seems compatible with relativistic space-time and does not seem to licence any superluminal causation.
\end{itemize}
Challenges and avenues for future work on the Q-based interpretation---partly discussed here, partly not---include (but are not restricted to) those that arise from the questions:
\begin{itemize}
	\item Is the Q-based interpretation really empirically adequate?
	\item How can Drummond's insights into Q-function dynamics and underlying microdynamics be generalized to quantum field theories beyond the massless bosonic case?
	\item Should the Q-based interpretation be combined with an epistemic or ontic account of the Q-function?
	\item When discussing foundational problems in quantum statistical mechanics that arise already in classical statistical mechanics---e.g. the problem of the arrow of time---what is the effect of considering the Q-function as the quantum analogue of the classical probability distribution $\rho$?
	\item Which philosophical accounts of probability, the laws of nature, time, causality and other key scientific concepts fit well with the Q-based interpretation?
\end{itemize}
The purpose of this paper has been fulfilled if readers are convinced that the Q-based interpretation is sufficiently promising that trying to answer these questions is worth some effort.

\section*{Acknowledgements}
I would like to thank Ken Wharton for drawing my attention to (\citet{drummond}), without which this work would not have happened. He, Peter Drummond, Pete Evans, and Ronnie Hermens gave very helpful feedback on earlier drafts of this article.  I am extremely grateful to two anonymous referees for incredibly constructive comments and suggestions.

\begin{flushright}
\textit{
Simon Friederich\\
University of Groningen\\
University College Groningen and Faculty of Philosophy\\
Groningen, The Netherlands\\
s.m.friederich@rug.nl
}
\end{flushright}


\begin{thebibliography}{*}
	

\bibitem[Adlam(2018)]{adlam} Adlam, E. [2018]: `Spooky action at a temporal distance', \textit{Entropy}, 20:41.
\bibitem[Bell(1964)]{belltheorem} Bell, J. S. [1964]: `On the Einstein-Podolsky-Rosen paradox', \textit{Physics}, 1:195-200, repr. in: J. S. Bell, \textit{Speakable and Unspeakable in Quantum Mechanics}, 2nd edition, 2004, pp.\ 14-21.
\bibitem[Bell(1982)]{bell} Bell, J. S. [1982]: `On the impossible pilot wave', \textit{Foundations of Physics}, 12:989-99, repr. in: J. S. Bell, \textit{Speakable and Unspeakable in Quantum Mechanics}, 2nd edition, 2004, pp.\ 159-68.
\bibitem[Colbeck and Renner(2017)]{colbeckrenner} Colbeck, R. and Renner, R. [2017]: `A system's wave function is uniquely determined by its underlying physical state', \textit{New Journal of Physics}, \textbf{19}, p. 013016. 
\bibitem[Colom\'{e}s, Zhan, and Oriols(2015)]{colomes} Colom\'{e}s, E., Zhan, Z and Oriols, X. [2015]: `Comparing Wigner, Husimi and Bohmian distributions: which one is a true probability distribution in phase space?', \textit{Journal of Computational Electronics}, \textbf{4}, pp.\ 894-906. 
\bibitem[Drummond(2021)]{drummond} Drummond, P. D. [2021]: `Time-evolution with symmetric stochastic action,' \textit{Physical Review Research}, \textbf{3}, p.\ 013240.
\bibitem[Drummond and Reid(2020)]{drummondreid} Drummond, P. D. and Reid, M. D. [2020]: `Retrocausal model of reality for quantum fields,' \textit{Physical Review Research}, \textbf{2}, p.\ 033266.
\bibitem[Evans(2015)]{evans} Evans, P. W. [2015]: `Retrocausality at no extra cost', \textit{Synthese}, \textbf{192}, pp.\ 1139-55.
\bibitem[Evans, Price and Wharton(2013)]{evansetal} Evans, P. W., Price, H. and Wharton, K. B. [2013]: `New slant on the EPR-Bell experiment', \textit{British Journal for the Philosophy of Science}, \textbf{64}, pp.\ 297-324.
\bibitem[Frauchiger and Renner(2018)]{frauchigerrenner} Frauchiger, D. and Renner, R. [2018]: `Quantum theory cannot consistently describe the use of itself', \textit{Nature Communications} \textbf{9}, p.\ 3711.
\bibitem[Friederich(2014)]{friederich} Friederich, S. [2014]: \textit{Interpreting Quantum Theory: A Therapeutic Apprach}, Houndmills, Basingstoke: Palgrave Macmillan.
\bibitem[Friederich and Evans(2019)]{friederichevans} Friederich, S. and Evans, P. W. [2019]: `Retrocausality in quantum mechanics', E. N. Zalta (ed.), \textit{The Stanford Encyclopedia of Philosophy}, Summer 2019 edition, URL = $<$https://plato.stanford.edu/archives/sum2019/entries/qm-retrocausality/$>$.
\bibitem[Fuchs and Schack(2013)]{fuchsschack} Fuchs, C. A. and Schack, R. [2013]: `Quantum-Bayesian coherence', \textit{Reviews of Modern Physics}, \textbf{85}, pp.\ 1693-715. 
\bibitem[Glauber(1963)]{glauber} Glauber, R. J. [1963]: `Coherent and incoherent states of the radiation field,' \textit{Physical Review}, \textbf{131}, p.\ 2766.  
\bibitem[Goldstein, Norsen, Tausk and Zanghi(2011)]{gntz} Goldstein, S., Norsen, T., Tausk, D. V. and Norsen, N.  [2011]: `Bell's theorem,' \textit{Scholarpedia}, \textbf{6}, p.\ 8378.
\bibitem[Hardy(2013)]{hardy} Hardy L. [2013]: `Are quantum states real?,' \textit{International Journal of Modern Physics B}, \textbf{27}, p.\ 1345012.
\bibitem[Harrigan and Spekkens(2010)]{harriganspekkens}  Harrigan, N. and Spekkens, R. W. [2010]: `Einstein, incompleteness, and the epistemic view of quantum states,' \textit{Foundations of Phyics}, \textbf{40}, pp.\ 125-57.
\bibitem[Healey(2017)]{healey} Healey, R. A. [2017]: \textit{The Quantum Revolution in Philosophy}, Oxford: Oxford University Press.
\bibitem[Husimi(1940)]{husimi} Husimi, K., [1940]: `Some formal properties of the density matrix,' \textit{Proceedings of the Physico-Mathematical Society of Japan}, \textbf{22}, p.\ 264.
\bibitem[Kirkwood(1933)]{kirkwood} Kirkwood, J. G. [1933]. `Quantum statistics of almost classical assemblies,' \textit{Physical Review}, \textbf{44}, p.\ 31.
\bibitem[Landsman(2007)]{landsman} Landsman, N. P. [2007]: `Between classical and quantum,' in: J. Butterfield and J. Earman (ed.), \textit{Philosophy of Physics: Handbook of the Philosophy of Science}, Elsevier, pp.\ 417-553.
\bibitem[Lazarovici and Hubert(2019)]{lazarovici} Lazarovici, D. and Hubert, M. [2019]: `How Quantum Mechanics can consistently describe the use of itself,' \textit{Scientific Reports}, \textbf{9}, p.\ 470.
\bibitem[Lee(1995)]{lee} Lee, H.-W. [1995]: `Theory and application of the quantum phase-space distribution functions', \textit{Physics Reports}, \textbf{259}, pp.\ 147-211.
\bibitem[Leifer(2014)]{leifer} Leifer, M. S. [2014]: `Is the quantum state real? An extended review of $\psi$-ontology theorems,' \textit{Quanta}, \textbf{3}, pp.\ 67-155. 
\bibitem[Leifer and Pusey(2017)]{leiferpusey} Leifer, M. S. and Pusey, M. F. [2017]: `Is a time symmetric interpretation of quantum theory possible without retrocausality?,' \textit{Proceedings of the Royal Society A}, \textbf{473}, p.\ 20160607.
\bibitem[Pusey, Barrett and Rudolph(2012)]{pbr} Pusey M. F., Barrett, J., and Rudolph, T. [2012]: `On the reality of the quantum state,' \textit{Nature Physics} \textbf{8}, pp.\ 475-78. 
\bibitem[Rosales-Z\'arate and Drummond(2015)]{Rosales-Zarate} Rosales-Z\'arate, L. E. C. and Drummond, P. [2015]: `Probabilistic Q-function distributions in fermionic phase-space', \textit{New Journal of Physics}, \textbf{17}, p.\ 032002.
\bibitem[Shrapnel and Costa(2018)]{shrapnelcosta} Shrapnel, S. and Costa, F. [2018]: `Causation does not explain contextuality,' \textit{Quantum}, \textbf{2}, p.\ 63. 
\bibitem[Spekkens(2005)]{spekkens} Spekkens, R. W. [2005]: `Contextuality for preparations, transformations, and unsharp measurements', \textit{Physical Review A}, \textbf{71}, p.\ 052108.
\bibitem[Spekkens(2007)]{spekkens_epistemic} Spekkens R. W. [2007]: `Evidence for the epistemic view of quantum states: a toy theory', \textit{Physical Review A}, \textbf{75}, p.\ 032110.
\bibitem[Sudarshan(1963)]{sudarshan} Sudarshan, E. C. G. [1963]: `Equivalence of semiclassical and quantum mechanical descriptions of statistical light beams,' \textit{Physical Review Letters}, \textbf{10}, p.\ 277.  
\bibitem[Wallace(2012)]{wallace} Wallace, D. [2012]: \textit{The Emergent Multiverse: Quantum Theory according to the Everett Interpretation}, Oxford: Oxford University Press.
\bibitem[Wharton(2016)]{wharton} Wharton, K. B. [2016]: `Towards a realistic parsing of the Feynman path integral', \textit{Quanta}, \textbf{5}, p.\ 1.
\bibitem[Wharton and Argaman(2020)]{whartonargaman} Wharton, K. B. and Argaman, N. [2020]: `\textit{Colloquium}: Bell's theorem and locally mediated reformulations of quantum mechanics', \textit{Reviews of Modern Physics}, \textbf{92}, p.\ 021002.
\bibitem[Wigner(1932)]{wigner} Wigner, E. [1932]: `On the quantum correction for thermodynamic equilibrium', \textit{Physical Review}, \textbf{40}, p.\ 749.

\end{thebibliography}
\end{document}